\documentclass[12pt,preprint]{aastex}
\def\jcap{JCAP}
\def\beq{\begin{equation}}
\def\eeq{\end{equation}}
\def\ben{\begin{eqnarray}}
\def\een{\end{eqnarray}}
\def\numass{\sum m_{\nu}}

\def\ev{\,{\rm eV}}
\def\munit{\,h^{-1}\! M_{\odot}}

\usepackage{color}

\begin{document}
\title{Breaking the Dark Degeneracy with the Drifting Coefficient of the Field Cluster Mass Function}
\author{Suho Ryu\altaffilmark{1}, Jounghun Lee\altaffilmark{1}, 
Marco Baldi\altaffilmark{2,3,4}}
\altaffiltext{1}{Astronomy Program, Department of Physics and Astronomy, FPRD, 
Seoul National University, Seoul 08826, Korea \email{shryu@astro.snu.ac.kr, jounghun@astro.snu.ac.kr}}
\altaffiltext{2}{Dipartimento di Fisica e Astronomia, Alma Mater Studiorum Universit\`a di Bologna, viale Berti Pichat, 
6/2, I-40127 Bologna, Italy}
\altaffiltext{3}{INAF - Osservatorio Astronomico di Bologna, via Ranzani 1, I-40127 Bologna, Italy}
\altaffiltext{4}{INFN - Sezione di Bologna, viale Berti Pichat 6/2, I-40127 Bologna, Italy}
\begin{abstract}
We present a numerical analysis supporting the evidence that the redshift evolution of the drifting coefficient of the field cluster mass function 
is capable of breaking several cosmic degeneracies. 
This evidence is based on the data from the {\small CoDECS} and {\small DUSTGRAIN}-{\it pathfinder} simulations performed separately for various 
non-standard cosmologies including coupled dark energy, $f(R)$ gravity and combinations of $f(R)$ gravity with massive neutrinos 
as well as for the standard $\Lambda$CDM cosmology. We first numerically determine the field cluster mass functions at various redshifts 
in the range of $0\le z\le 1$ for each cosmology. Then, we compare the analytic formula developed in previous works with the numerically 
obtained field cluster mass functions by adjusting its drifting coefficient, $\beta$, at each redshift. It is found that the analytic formula 
with the best-fit coefficient provides a good match to the numerical results at all redshifts for all of the cosmologies. The empirically 
determined redshift evolution of the drifting coefficient, $\beta(z)$, turns out to significantly differ among different cosmologies. 
It is also shown that even without using any prior information on the background cosmology the drifting coefficient, $\beta(z)$, can 
discriminate with high statistical significance the degenerate non-standard cosmologies not only from the $\Lambda$CDM but 
also from one another. It is concluded that the evolution of the departure from the Einstein-de Sitter state and spherically symmetric collapse 
processes quantified by $\beta(z)$ is a powerful probe of gravity and dark sector physics. 
\end{abstract}
\section{Introduction}

A cosmic degeneracy refers to the circumstance that a standard diagnostic fails to distinguish between different cosmologies with high statistical significance. 
For example, the cluster mass function, which is regarded as one of the most powerful probes of cosmology based on the large 
scale structure, is unable to discriminate the effect of a low amplitude of the linear density power spectrum from that of massive neutrinos ($\nu$) 
(dubbed the $\sigma_{8}$-$\sum m_{\nu}$ degeneracy) in a $\nu\Lambda$CDM (massive neutrinos $\nu$ + cosmological constant $\Lambda$ + Cold Dark Matter) 
cosmology. 
Another example that the cluster mass function fails to discriminate from the $\Lambda$CDM cosmology is a coupled dark energy (cDE) model in which a scalar field 
DE coupled to DM particles follows a supergravity potential \citep{bal-etal10}. 
Since the cosmic degeneracy is caused by the limited sensitivity of a given standard diagnostic on which the degenerate models have almost the same effects, 
what is required to break it is to overcome the limitation by utilizing prior information from other independent diagnostics. In the aforementioned example, 
the $\sigma_{8}$-$\sum m_{\nu}$ degeneracy can be broken by prior information on the large-scale amplitude of the linear density power spectrum from the CMB 
observations.

There are, however, a few cosmic degeneracies which have been found more difficult to break even by combining the priors from several independent diagnostics. 
A notorious example is the cosmic degeneracy between the $\Lambda$CDM + GR and the $\nu$CDM + MG cosmologies, where GR and MG stand for the 
general relativity and modified gravity, respectively \citep[see e.g.][]{bal-etal14,wri-etal19}. All different versions of the MG theory adopt a common tenet that the apparent acceleration of the present Universe 
is caused not by the dominance of the anti-gravitational $\Lambda$ at the present epoch but by the deviation of the gravitational law from the prediction of GR on cosmological scales. The consequence of this tenet is the existence of a long-range fifth force, which in turn has an effect of enhancing the density power spectrum 
on the scales comparable to those affected by the suppression due to free streaming massive neutrinos \citep[for a review, see][]{mgreview}. 

In the theory of $f(R)$ gravity, the gravitational dynamics is defined by a modified Einstein-Hilbert action functional to which 
an arbitrary function of the Ricci scalar, $f(R)$, is introduced as a substitution for the Ricci scalar $R$ itself of the original action in GR \citep[see e.g.,][]{buc70,sta80,HS07}. 
Choosing as a viable MG the $f(R)$ gravity in which the Ricci scalar term, $R$, in the Einstein-Hilbert action functional is replaced by an arbitrary function, 
$f(R)$, \citet{bal-etal14} numerically investigated a possible cosmic degeneracy between the $\Lambda$CDM + GR and the $\nu$CDM + $f(R)$, 
and demonstrated that the two cosmologies cannot be discriminated from each other by several standard diagnostics such as the nonlinear density power spectra, 
halo bias and cluster mass functions \citep[see also][]{hag-etal19,gf-etal19}. The nonlinear growth rate functions, cluster velocity dispersions, and tomographic higher-order weak lensing statistics were proposed 
in subsequent works as candidate diagnostics that could be capable of breaking this cosmic degeneracy \citep{gio-etal19,pel-etal18,hag-etal19}, also employing Machine Learning techniques \citep[][]{pel-etal19,mer-etal19}.

Very recently, \citet{RL20a} have developed a new independent diagnostics based on the evolution of the drifting coefficient of the field cluster abundance and shown that 
this new diagnostics is capable of distinguishing between those dynamical DE cosmologies which are degenerate with the $\Lambda$CDM case in their linear density 
power spectra and cluster mass functions. In the follow-up work of \citet{RL20b}, it was also found that the aforementioned $\sigma_{8}$-$\sum m_{\nu}$ degeneracy can 
be in principle broken by this new diagnostics. Our goal here is to explore whether or not this new diagnostics can break other cosmic degeneracies including that 
between the $\Lambda$CDM + GR and the $\nu$CDM + MG models. 

The key contents of the upcoming Sections are as follows. Section \ref{sec:review} will present a succinct review of the analytic model for the field cluster abundance on which 
the new diagnostics is based. Section \ref{sec:cde} will present a numerical evidence that the new diagnostics can distinguish between the standard 
$\Lambda$CDM and a cDE cosmology that produce very similar cluster mass functions. Section \ref{sec:fr} will present a proof for the validity of the analytic formulae for 
the field mass function and drifting coefficient in a MG gravity cosmology. Section \ref{sec:frnu} will present a numerical evidence that the new diagnostics can break 
the degeneracy between the standard $\Lambda$CDM+GR and the $\nu$CDM+MG cosmologies. Section \ref{sec:con} will be devoted to summarizing the final results and 
discussing the follow-up works.

\section{A Succinct Review of the Analytic Model}\label{sec:review}

According to the generalized excursion set theory that incorporates the non-spherical collapse conditions and non-Markovian random-walk 
process \citep{CA11a,CA11b}, the differential mass function of the cluster halos can be written as:
\begin{equation}
\label{eqn;exc}
\frac{d\,N(M, z)}{d\,{\rm ln}\,M} = \frac{\bar{\rho}_{m}}{M}\Bigg{\vert}\frac{d\,{\rm ln}\,\sigma^{-1}}{d\,{\rm ln}\,M}\Bigg{\vert}f[\sigma(M, z)]\:,
\label{eqn:exc}
\end{equation}
where $\bar{\rho}_{m}$ is the mean mass density of the universe, $\sigma (M)$ is the standard deviation of the linear density contrast field 
smoothed on the mass scale $M$, and $f(\sigma)$ is the multiplicity function characterized by two coefficients, $\beta$ and $D_{B}$: 
\begin{eqnarray}
\label{eqn:multi}
f(\sigma ; D_{B},\beta) &\approx& f^{(0)}(\sigma ; D_{B}, \beta) + 
f^{(1)}_{\beta=0}(\sigma ; D_{B}) + 
f^{(1)}_{\beta}(\sigma ; D_{B}, \beta) + 
f^{(1)}_{\beta^2}(\sigma ; D_{B}, \beta)\, , \\
\label{eqn:f0}
f^{(0)}(\sigma ; D_{B}, \beta) &=& \frac{\delta_{sc}}{\sigma\sqrt{1+D_{B}}} \sqrt{\frac{2}{\pi}}\,
e^{-\frac{(\delta_{sc}+\beta \sigma^2)^2}{2\sigma^2(1+D_{B})}}\, ,\\
\label{eqn:f1b0}
f^{(1)}_{\beta=0}(\sigma ; D_{B}) &=&-\tilde{\kappa}\frac{\delta_{sc}}{\sigma}
\sqrt{\frac{2a}{\pi}}\left[e^{-\frac{a\delta_{sc}^2}{2\sigma^2}}
-\frac{1}{2}\Gamma\left(0,\frac{a \delta_{sc}^2}{2 \sigma^2}\right)\right]\, , \\
\label{eqn:f1b1}
f^{(1)}_{\beta}(\sigma ; D_{B}, \beta) &=&
-\beta\,a\,\delta_{sc}\left[f^{(1)}_{\beta=0}(\sigma; D_B)+\tilde{\kappa}\,
\textrm{erfc}\left(\frac{\delta_{sc}}{\sigma}\sqrt{\frac{a}{2}}\right)\right]\, , \\
\label{eqn:f1b2}
f^{(1)}_{\beta^2}(\sigma ; D_{B}, \beta) &=&\beta^{2}a^{2}\delta^{2}_{sc}\tilde{\kappa}
\biggl\{\textrm{erfc}\left(\frac{\delta_{sc}}{\sigma}\sqrt{\frac{a}{2}}\right)+\\
&& \frac{\sigma}{a\delta_{sc}}\sqrt{\frac{a}{2\pi}}\biggl[e^{-\frac{a\delta_{sc}^2}
{2\sigma^2}}\left(\frac{1}{2}-\frac{a \delta_{sc}^2}{\sigma^2}\right)+\frac{3}{4}\frac{a\delta_{sc}^2}
{\sigma^2}\Gamma\left(0,\frac{a \delta_{sc}^2}{2 \sigma^2}\right)\biggr]\biggr\}\, ,
\end{eqnarray}
with $a\equiv 1/(1+D_B)$, $\tilde{\kappa} = \kappa a$, $\kappa = 0.475$, upper incomplete gamma function $\Gamma(0, x)$, complementary 
error function ${\rm erfc}(x)$, and critical density contrast for the spherical collapse $\delta_{sc}$. 
The diffusion coefficient, $D_{B}$, is a measure of the stochasticity of $\delta_{c}$ caused by the influence of the perturbing neighbors as well as by the 
uncertainty in the identification of a cluster halo, while the drifting coefficient, $\beta$, quantifies the deviation of the critical density contrast for the realistic non-spherical 
collapse, $\delta_{c}$, from $\delta_{sc}$ at a given $\sigma (M)$.

Suggesting that for the isolated field clusters, the degree of the stochasticity of $\delta_{c}$ should be negligible, (i.e., $D_{B}=0$), \citet{lee12} modified the 
Corasaniti-Achitouv formalism to construct a single parameter model for the field cluster mass function as 
\begin{equation}
\label{eqn:exc_f}
\frac{dN_{\rm I}(M, z)}{d\ln M} = \frac{\bar{\rho}}{M}\Bigg{\vert}\frac{d\,{\rm ln}\,\sigma^{-1}}{d\,{\rm ln}\,M}\Bigg{\vert}
f\left[\sigma(M,z) ; D_{B}=0,\beta\right]\, .
\end{equation}
In spite of having only single parameter, this analytic model has been shown to be very successful in describing the field cluster 
mass functions in a wide redshift range not only for the standard $\Lambda$CDM cosmology but also for several non-standard 
cosmologies including dynamical dark energy or massive neutrinos \citep{RL20a,RL20b}.

Although the exact value of $\delta_{sc}$ has been known to weakly depend on the background cosmology as well as on the redshift 
\citep{eke-etal96,pac-etal10}, \citet{RL20a} regarded $\delta_{sc}$ as a constant, setting it at the Einstein-de Sitter value of $1.686$ \citep{GG72}, 
as done in the original formulation of the generalized excursion set mass function theory \citep{MR10a,MR10b}. 
In reality, the gravitational collapse proceeds in a non-spherical way, for which the actual critical density contrast, $\delta_{c}$, 
departs from the idealistic spherical threshold, $\delta_{sc}$. The cosmology dependence of $\delta_{c}$ is expected to overwhelm 
that of $\delta_{sc}$, given that the degree of the non-sphericity of the collapse process is closely linked with the anisotropy of the cosmic web, which in turn 
possesses strong dependence on the background cosmology \citep[e.g.,][]{SL13,nai-etal20}. 
Unlike $\delta_{sc}$, however, the value of $\delta_{c}$ and its link to the initial conditions cannot be analytically derived from first principles 
due to the complexity associated with the non-spherical collapse process \citep{BM96}. 

\citet{RL20a} imparted any redshift and cosmology dependences of the collapse threshold at a given mass scale to the empirical parameter, $\beta$, 
and showed that the redshift evolution of the empirically determined $\beta(z)$ has a universal form of an inverse sine hyperbolic function of $z$, 
regardless of the background cosmology: 
\begin{equation}
\beta(z) = \beta_{A}\ {\sinh}^{-1} \left[\frac{1}{q_z}(z-z_c)\right]\, ,
\label{eqn:inv_sinh} 
\end{equation}
where $\beta_{A}$, $q_{z}$ and $z_{c}$, represent three adjustable parameters, whose best-fit values depend on the background cosmology. 
Especially, the critical redshift, $z_{c}$, defined as $\beta(z_{c})=0$ (i.e., $\delta_{c}(z_{c})=\delta_{sc}=1.686$), has been found to depend most 
sensitively on the background cosmology \citep{RL20a,RL20b}, as it reflects not only how severely the real gravitational collapse deviates from the 
spherical symmetry but also how rapidly the universe evolves away from the Einstein-de Sitter state.

\section{Evolution of the Drifting Coefficient in Non-Standard Cosmologies}

\subsection{Effect of Coupled Dark Energy on $\beta(z)$}\label{sec:cde}

A cDE cosmology describes an alternative universe where the role of DE is played by a dynamical scalar field, $\phi$, coupled to DM particles through energy-momentum exchange. The DE-DM coupling that causes the time-variation of DM particle mass \citep{wet95,ame00,ame04}
generates a long-range fifth force via which the growth of structures can be enhanced \citep[e.g.,][and references therein]{man-etal03,mac-etal04,MB06,PB08,bal-etal10,WP10}.
Categorized by the shape of DE self-interaction potential, $V(\phi)$, as well as by the strength of the DE-DM coupling, $s(\phi)\equiv -d\ln m_{\rm DM}/d\phi$, 
a cDE cosmology has recently attained delving attentions since it has been found to provide a possible solution to the Hubble tension \citep{cde_hubble20}.

To investigate the effect of cDE on the redshift evolution of $\beta(z)$, we utilize the data from the Large Coupled Dark Energy Cosmological 
Simulations ({\small L-CoDECS}) run by \citet{codecs1} with a modified version of the {\small GADGET3} code, a non-public developers version of the widely-used public code 
GADGET-2 \citep[][]{Gadget2}. The {\small L-CoDECS} is a series of $N$-body cosmological runs that simulate a standard $\Lambda$CDM and five different cDE cosmologies 
on a periodic box of linear size $1\, h^{-1}$Gpc containing $1024^{3}$ collisionless DM particles of 
individual mass $m_{\rm DM}=5.84\times 10^{10}\munit$ as well as an equal number of collisionless baryon particles of $m_{\rm baryon}=1.17\times 10^{10}\munit$. 
The initial conditions of the standard $\Lambda$CDM cosmology were chosen to meet the constraints from the Seven-Year Wilkinson Microwave Anisotropy 
observations \citep{wmap7}.
The five different cDE cosmologies are divided into three categories: the constant DM-DE coupling and exponential potentials (EXP001, EXP002, EXP003), 
the exponential DM-DE coupling and exponential potential (EXP008e3) and the constant coupling and supergravity potential (SUGRA). 
All cDE cosmologies simulated by the L-CoDECS were ensured to have a flat geometry, sharing the same values of the five key cosmological parameters, 
$h = 0.703$, $\Omega_{\rm CDM} = 0.226$, $\Omega_{\rm DE} = 0.729$, $\Omega_b = 0.0451$, $A_s = 2.42 \times 10^{-9}$ and $n_s = 0.966$. 
They differ from one another only in the potential shape and DM-DE coupling as well as in the linear density power spectrum amplitude, information on which 
are provided in the first four columns of Table \ref{tab:mds_cde}. 
For more detailed description of the cDE cosmologies and the L-CoDECS \footnote{All data are available at the CoDECS website, http://www.marcobaldi.it/web/CoDECS.html}, 
we refer the readers to \citet{codecs1,codecs2}.

The {\small L-CoDECS} simulations have been released with catalogs of gravitationally bound halos identified for each cosmology through a two-step process starting with a 
Friends-of-Friends (FoF) algorithm with linking length parameter of $l_{c}=0.2$ followed by a gravitational unbinding procedure of each individual FoF halo using the 
{\small SUBFIND} algorithm \citep[][]{spr-etal01} that allows to associate spherical overdensity masses and radii to each gravitationally bound main substructure. Selecting the 
cluster halos with virial masses $M\ge 3\times 10^{13}\munit$ from the halo catalog and applying again the FoF algorithm with linking length parameter of $l_{sc}=2\times l_{c}$ 
to such halo sample, we identify the clusters of cluster halos as marginally bound superclusters. Here, the choice of $l_{sc}=2\times l_{c}$ for the FoF finding of the superclusters 
is made to guarantee $D_{B}=0$ for the field cluster halos, as explained in \citet{RL20a}.

Sorting out the field cluster halos as the superclusters which have only one member cluster halo, we create 
a sample of the field cluster halos at each redshift in the range of $0\le z\le 1$. The differential mass function of the field clusters halos is determined by 
computing the number density of the field cluster halos, $dN_{\rm I}/d\ln M$, whose masses fall in the differential bin of the logarithmic mass, $[\ln M, \ln M + d\ln M]$, 
at each redshift. As done in \citet{RL20a}, we employ the Jackknife method to compute the errors in the determination of $dN_{\rm I}/d\ln M$. Splitting the sample of the field clusters 
into $8$ Jackknife subsamples of equal size, we obtain $dN_{\rm I}/d\ln M$ separately from each subsample and then calculate the one standard deviation scatter of $dN_{\rm I}/d\ln M$ 
among the $8$ subsamples as errors at each logarithmic mass bin. 

Using the linear density power spectrum of each cDE cosmology provided within the {\small CoDECS} public data release, we evaluate the linear density rms fluctuation, 
$\sigma(M)$, and the analytic mass function of the field cluster, Equation (\ref{eqn:exc_f}), as well. 
The best-fit value of the drifting coefficient, $\beta$ in Equation (\ref{eqn:exc_f}) is determined at each redshift by minimizing the following $\chi^{2}$ under the assumption 
that there is no correlation in $dN_{\rm I}/d\ln M$ at different mass bins:
\begin{equation}
\chi^{2}(\beta)=\sum_{i=1}^{N_{p}}\frac{\left[n^{\rm n}(\ln M_{i})-n^{\rm a}(\ln M_{i};\beta)\right]^{2}}{\sigma^{2}_{n_{i}}} \, ,
\end{equation} 
where $n^{\rm n}(\ln M_{i})$ and $n^{\rm a}(\ln M_{i};\beta)$ are the numerically obtained and analytically evaluated values of 
$dN_{\rm I}/d\ln M$ at the $i$th logarithmic mass bin, respectively,  $N_{\rm p}$ is the number of the logarithmic mass bins and $\sigma_{n_{i}}$ is the Jackknife error 
on $n^{\rm n}(\ln M_{i})$. The error on $\beta$ is then determined as the square root of the inverse of the second derivative of $\chi^{2}$ with respect to $\beta$ as 
$\sigma_{\beta}=(d^{2}\chi^{2}/d\beta^{2})^{-1/2}$ \citep{RL20a}.

As done in \citet{RL20a}, once the values of $\beta(z)$ are determined at various redshifts, we fit them to Equation (\ref{eqn:inv_sinh}) via a nonlinear least square 
regression procedure with the SciPy python code \citep{vir-etal20} to find the best-fit values of the three parameters, $\beta_{A},\ q_{z}$ and $z_{c}$ and their associated 
errors $\sigma_{\beta_{A}},\ \sigma_{q_{z}}$ and $\sigma_{z_{c}}$, respectively (see Table \ref{tab:mds_cde}).  
Then, we calculate the statistical significance of the differences in the three parameters among the cosmologies as 
$\Delta\beta_{A}/\sigma_{\Delta\beta_{A}}$, $\Delta q_{z}/\sigma_{\Delta q_{z}}$ and $\Delta z_{c}/\sigma_{\Delta z_{c}}$ where 
$\Delta\beta_{A}$, $\Delta q_{z}$ and $\Delta z_{c}$ are the differences in the three parameters between two cosmologies, while 
$\sigma_{\Delta\beta_{A}}$, $\sigma_{\Delta q_{z}}$ and $\sigma_{\Delta z_{c}}$ correspond to the propagated errors in the determination of the differences. 

\begin{figure}[ht]
\begin{center}
\plotone{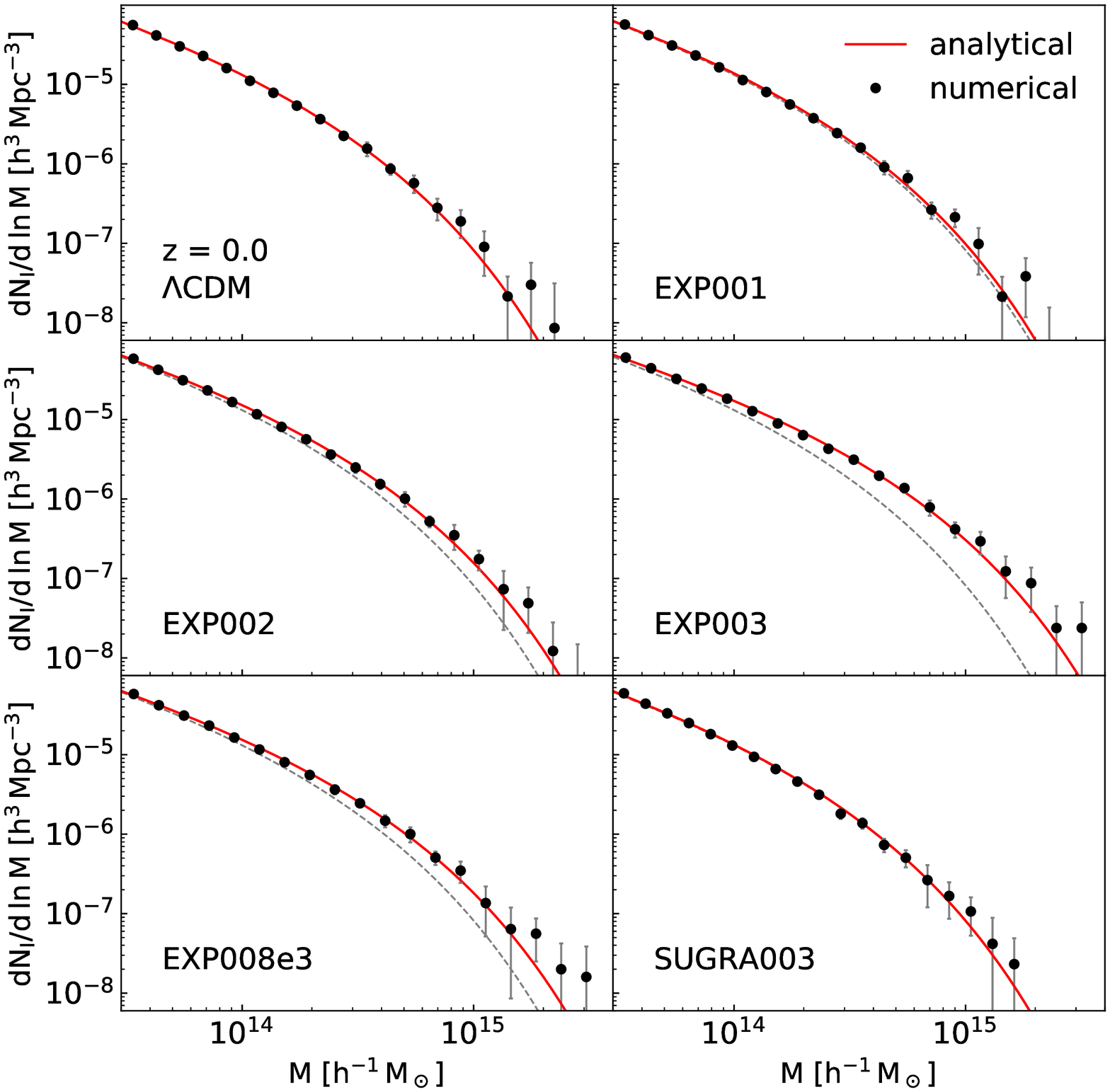}
\caption{Field cluster mass functions numerically obtained (black filled circles) from the CoDECS and analytic model with 
the best-fit drifting coefficient (red solid lines) for a $\Lambda$CDM and five different cDE cosmologies at $z=0$.}
\label{fig:mf_cde_z0}
\end{center}
\end{figure}

\begin{figure}[ht]
\begin{center}
\plotone{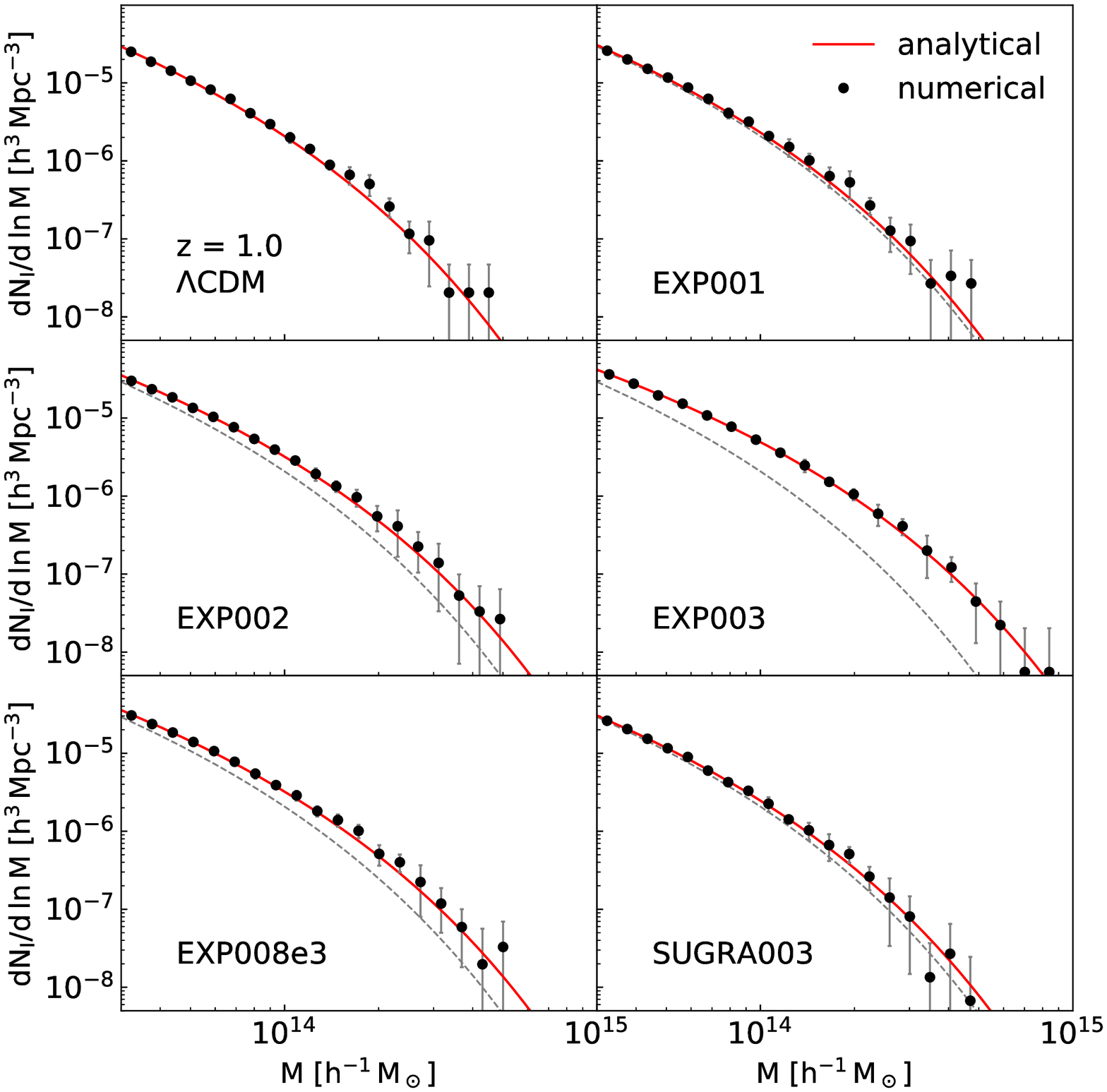}
\caption{Same as Figure \ref{fig:mf_cde_z0} but at $z=1$.}
\label{fig:mf_cde_z1}
\end{center}
\end{figure}

Figure \ref{fig:mf_cde_z0} (Figure \ref{fig:mf_cde_z1}) plots the numerically determined mass functions of the field cluster halos (filled black circles) as well as the analytic model 
(red solid line), Equation (\ref{eqn:exc_f}), with the best-fit value of $\beta$ for the six cosmologies at $z=0$ ($z=1$), respectively. In each panel, the analytic model for the 
$\Lambda$CDM case (black dashed line) is also plotted to clearly show the differences. Although the analytic model, Equation (\ref{eqn:exc_f}), succeeds in matching the numerical results at both of the redshifts for all of the cDE cosmologies, the field cluster mass functions are found to be incapable of telling apart with high statistical significance 
the three cosmologies $\Lambda$CDM, EXP001 and SUGRA at both of the redshifts, $z=0$ and $1$.

\begin{figure}[ht]
\begin{center}
\plotone{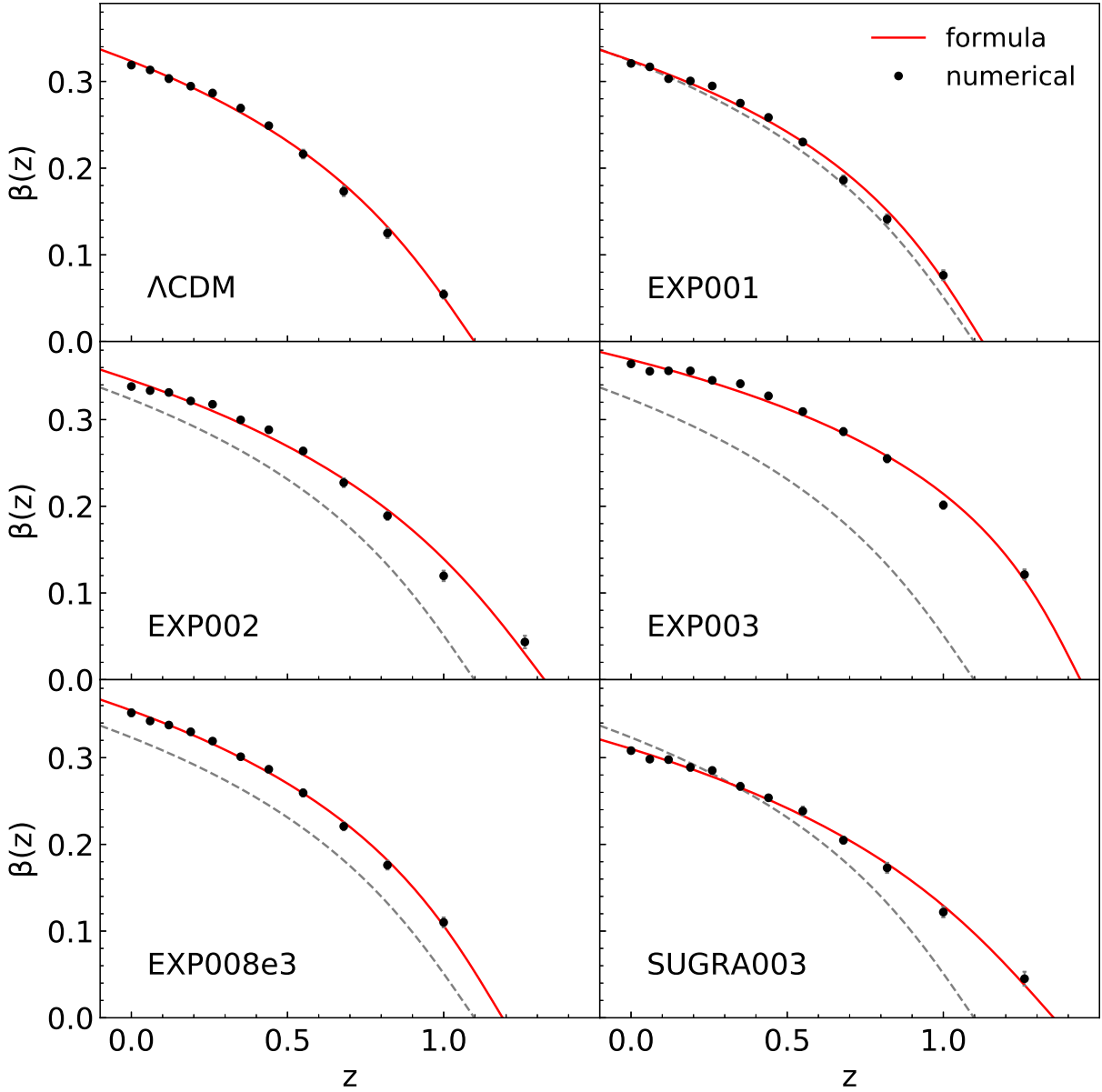}
\caption{Empirically determined redshift evolution of the drifting coefficient of the field clusters (black filled circles) and 
fitting formula (red solid lines) for six different cosmologies.}
\label{fig:beta_cde}
\end{center}
\end{figure}

Figure \ref{fig:beta_cde} plots the redshift evolution of the empirically determined drifting coefficient, $\beta(z)$ (filled black circles) as well as the fitting formula 
(red solid lines) for the six cosmologies. In each panel, the fitting formula for the $\Lambda$CDM case (black dashed line) are also plotted to show the differences. As can be seen, 
the fitting formula expressed in terms of the inverse sine hyperbolic function, Equation (\ref{eqn:inv_sinh}), with the best-fit values of $q_{z}$, $\beta_{A}$ and $z_{c}$ indeed describes 
quite well the behaviors $\beta(z)$ for all of the six cosmologies. Note that the SUGRA can be distinguished by $\beta(z)$ from the $\Lambda$CDM with high statistical 
significance despite that the two cosmologies are mutually degenerate in the cluster mass functions. The statistical significance of the difference in the critical redshift parameter, 
$z_{c}$, between the $\Lambda$CDM and the SUGRA cosmologies is found to be as high as $7.48\, \sigma $.
Although $\beta(z)$ distinguishes with high statistical significance the other cDE cosmologies except for the EXP001 from the $\Lambda$CDM, 
it fails to break the degeneracy between the $\Lambda$CDM and the EXP001 cases, due to the extremely weak DM-DE coupling of the latter cosmology. 

We have so far used prior information on the background cosmology for the determination of $dN_{\rm I}/d\ln M$ and $\beta(z)$. In other words, to examine if $\beta(z)$ can break a 
cosmic degeneracy between two different cosmologies, we assume that information on the shape of the linear density power spectrum are available. 
In practice, however, this prior information is not available for the determination of $\beta(z)$. Especially, if a background cosmology is indistinguishable from the $\Lambda$CDM 
case by the standard diagnostics, then it may not be justified to make such a preemptive assumption about the shape of the linear density power spectrum. 
The EXP001 corresponds to this case where no prior information on the background cosmology should be assumed to be available in practice, since the standard diagnostics 
including the linear density power spectrum, mass function, etc., are unable to distinguish it from the $\Lambda$CDM case. 

\begin{figure}[ht]
\begin{center}
\plotone{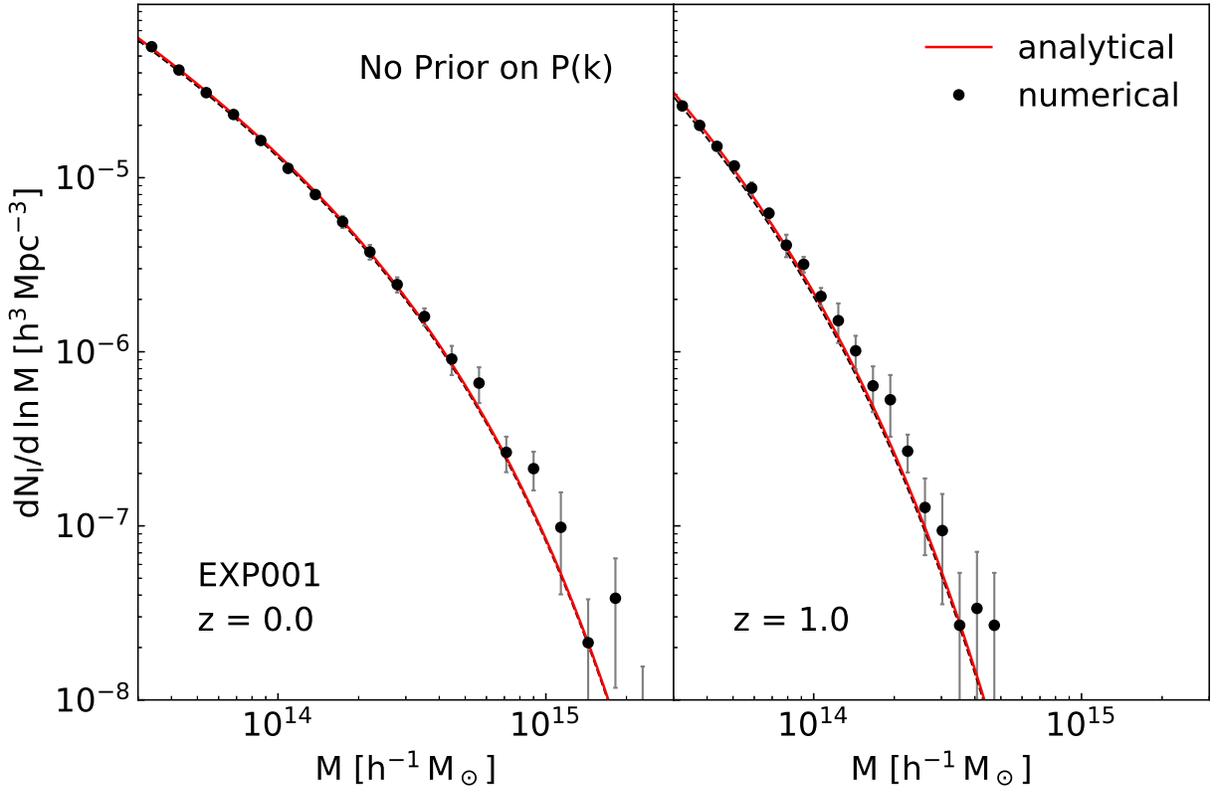}
\caption{Field cluster mass functions for the EXP001 case determined without using prior information on $P(k)$.}
\label{fig:noprior_mf_cde}
\end{center}
\end{figure}

To deal with this degeneracy, we use the linear density power spectrum of the $\Lambda$CDM case, $P(k;\Lambda{\rm CDM})$, for the computation of $\sigma(M)$ in 
$dN_{\rm I}/d\ln M$ for the EXP001 case and compare the reevaluated analytic model with the numerical results to find the best-fit $\beta(z)$. That is, 
we redetermine $\beta(z)$ for the EXP001 case without using prior information on $P(k;{\rm EXP001})$. 
Figure \ref{fig:noprior_mf_cde} plots the analytical mass function of the field clusters (red solid lines) obtained by using $P(k;\Lambda{\rm CDM})$ and compares it with the numerical results (black filled circles) for the EXP001 case at $z=0$ (top panel) and $z=1$ (bottom panel). As can be seen, in spite of no prior information on the background cosmology, the analytical mass function of the field clusters still describes quite well the 
numerical results at both of the redshifts for the EXP001 case. 

\begin{figure}[ht]
\begin{center}
\plotone{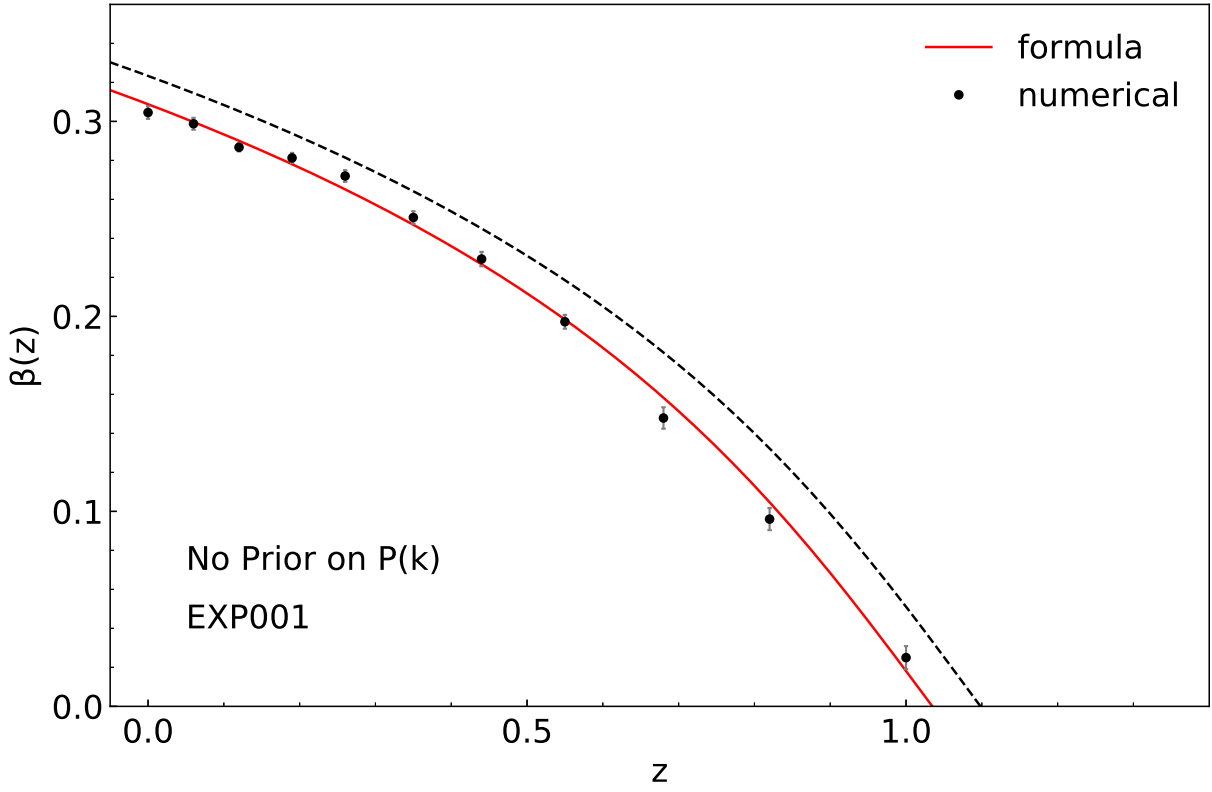}
\caption{Evolution of the drifting coefficient of the field clusters for the EXP001 case determined without using prior information on $P(k)$}
\label{fig:noprior_beta_cde}
\end{center}
\end{figure}

\begin{figure}[ht]
\begin{center}
\plotone{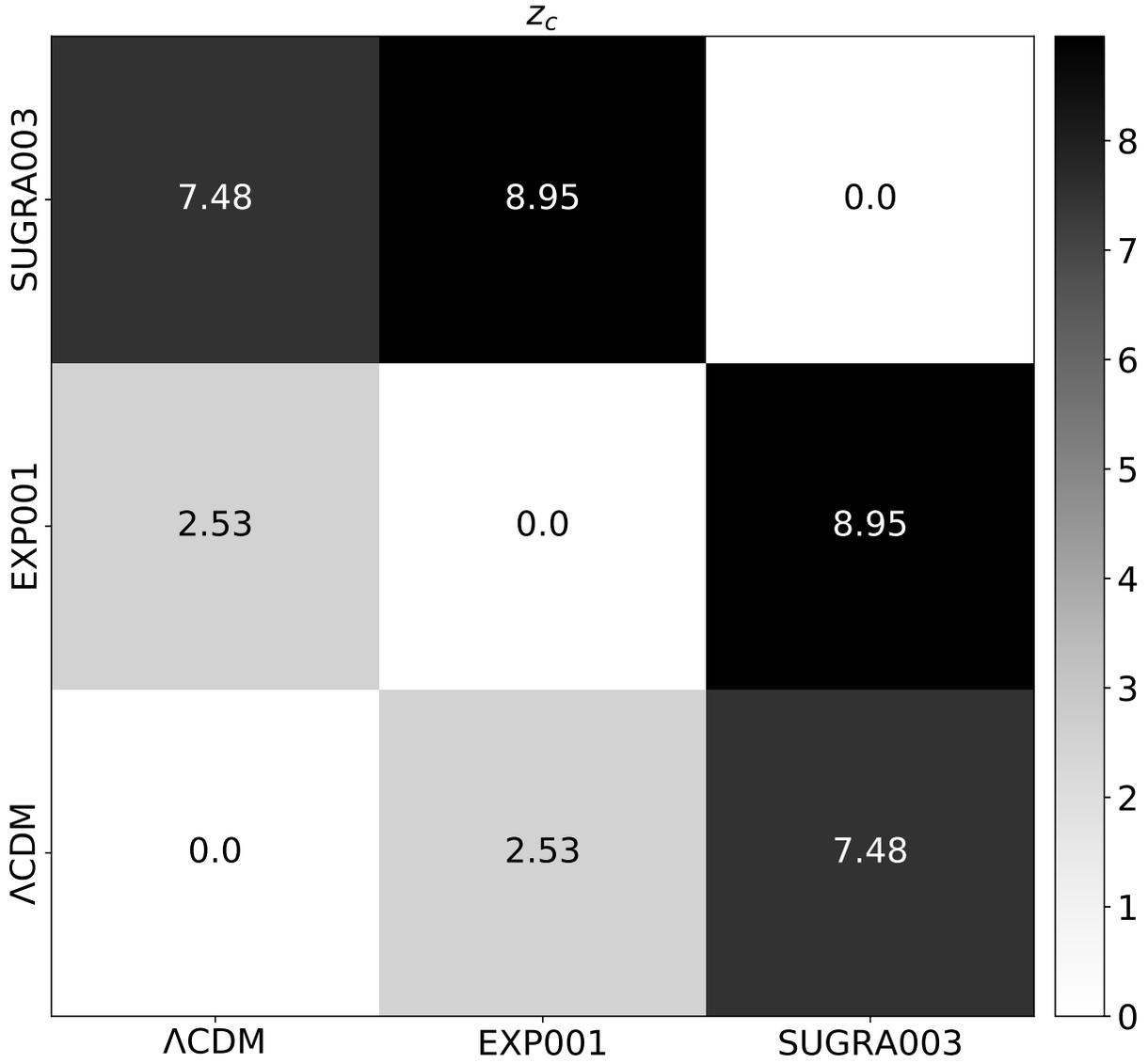}
\caption{Statistical significances of the differences among the $\Lambda$CDM, EXP001, and SUGRA that 
are mutually degenerate in the cluster mass functions.}
\label{fig:signi_cde}
\end{center}
\end{figure}

Figure \ref{fig:noprior_beta_cde} plots the same as the top-right panel of Figure \ref{fig:beta_cde} but without using prior information on $P(k;{\rm EXP001})$. 
As can be seen, the EXP001 turns out to yield larger differences in $\beta(z)$ from the $\Lambda$CDM. The best-fit values of $\beta_{A}$, $q_{z}$ and $z_{c}$ for the EXP001 case listed in Table \ref{tab:mds_cde} correspond to the ones obtained without using prior information 
on $P(k;{\rm EXP001})$. The statistical significance of the difference in $z_{c}$ between the $\Lambda$CDM and the EXP001 is found to be as high as $2.53$. 
Figure \ref{fig:signi_cde} summarizes the statistical significance of the difference in $z_{c}$ among the three cosmologies, $\Lambda$CDM, EXP001, and SUGRA, 
which are mutually degenerate in the field cluster mass functions. Although the degeneracy between the $\Lambda$CDM and the EXP001 can be broken by 
$\beta(z)$ only with $2.53$ significance, we speculate that a larger data set would improve the significance. 

\subsection{Effect of $f(R)$ Gravity on $\beta(z)$}\label{sec:fr}

In the theory of $f(R)$ gravity, the strength of a long range fifth force is quantified by the absolute value of the derivative of $f(R)$ with respect to the 
Ricci scalar $R$ at the present epoch, $\vert f_{R0}\vert\equiv \vert df/dR\vert_{0}$. A larger value of $\vert f_{R0}\vert$ corresponds to a stronger fifth force, 
which would more severely enhance the small-scale density power \citep{HS07,LB07}. 
If neutrinos have a non-zero mass in a $f(R)$ gravity cosmology, however, the suppressing effect of the free streaming neutrinos on the small-scale density power spectrum 
could compensate the enhancing effect of the fifth force, resulting in a suppression of the deviations from the standard $\Lambda$CDM + GR cosmology that each of these two scenarios would individually imprint on structure formation. In other words, the linear density power spectra may not be capable of distinguishing a certain combination of 
$f_{R0}$ with $\sum m_{\nu}$ from the standard $\Lambda$CDM + GR cosmology, since they could have zero net effect on the amplitude of small-scale density perturbations \citep[e.g.,][]{bal-etal14}. 

To investigate if $\beta(z)$ can also break the cosmic degeneracy between $\Lambda$CDM+GR and $\nu$CDM+$f(R)$, we use a subset of the data from 
the {\small DUSTGRAIN}-{\it pathfinder} $N$-body simulation suite that were conducted by \citet{gio-etal19} on a box of volume $750^{3}\,h^{-3}$Mpc$^{3}$ 
for various $\nu$CDM+$f(R)$ cosmologies as well as the $\Lambda$CDM+GR cosmology. The {\small DUSTGRAIN}-{\it pathfinder} simulations were performed with the {\small MG-GADGET} code \citep{mggadget} -- another modified version of {\small GADGET-3} implementing an adaptive mesh solver for the spatial fluctuations of the $f_{R}$ scalar degree of freedom -- to trace the evolution of 
$768^{3}$ DM particles of mass $8.1\times 10^{10}\,h^{-1}M_{\odot}$. To simulate the $\nu$CDM+$f(R)$ cosmologies, the 
{\small DUSTGRAIN}-{\it pathfinder} adopted the widely-used realisation of $f(R)$ proposed by \citet{HS07} and a particle-based implementation of massive neutrinos developed by \citet{vie-etal10}. 
Collapsed structures were identified through a FoF finder with a linking length parameter of $l_{c}=0.16$ followed by the unbinding procedure implemented in the {\small SUBFIND} code to identify the halo center and its spherical overdensity mass and radius for all gravitationally bound objects in each cosmology, similarly to what described above for the {\small CoDECS} simulations. For a detailed description of the technical details of the {\small DUSTGRAIN}-{\it pathfinder} simulations, 
see \citet{gio-etal19}. 

Among the various cosmologies simulated by the {\small DUSTGRAIN}-{\it pathfinder}, we consider three different CDM+$f(R)$ (namely, fR4, fR5 and fR6 corresponding to 
$\vert f_{R0}\vert=10^{-4},\ 10^{-5}$ and $10^{-6}$, respectively) and three different $\nu$CDM+$f(R)$ (namely fR4+$0.3\ev$, fR5+$0.15\ev$ and fR6+$0.06\ev$ 
corresponding to $\numass=0.3\ev,\ 0.15$ and $0.06\ev$, respectively) as well as the standard $\Lambda$CDM + GR (from here on, GR) with initial conditions 
set at the Planck values \citep{planck16}. These $7$ different cosmologies were ensured to be flat, described by the common key cosmological parameter values of 
$h= 0.67$, $\Omega_{m} = 0.31$, $\Omega_{\rm DE} = 0.67$, $\Omega_b = 0.0481$, $A_s = 2.2 \times 10^{-9}$ and $n_s = 0.97$. 
The first four columns of Table \ref{tab:mds_frnu} list the values of $\vert f_{R0}\vert$, $\sum m_{\nu}$, $\sigma_{8}$ for 
each of the seven cosmologies considered in the present work.

We first examine whether or not the analytic model for the field cluster mass function, Equation (\ref{eqn:exc_f}), is valid for the three CDM+$f(R)$ cosmologies. 
Analyzing the FoF halo catalogs in the redshift range $0\le z\le 1$ and following the same procedure described in Section \ref{sec:cde}, we numerically determine 
$dN_{\rm I}/d\ln M$ for the GR, fR4, fR5 and fR6 cases. 
To evaluate the analytic model, Equation (\ref{eqn:exc_f}), and compare it with the numerically determined $dN_{\rm I}/d\ln M$ to derive $\beta(z)$ for 
each of the three $f(R)$ gravity cosmologies, we use the {\small MGCAMB} code \citep{zha-etal09,hoj-etal11,zuc-etal19,camb}.
 
\begin{figure}[ht]
\begin{center}
\plotone{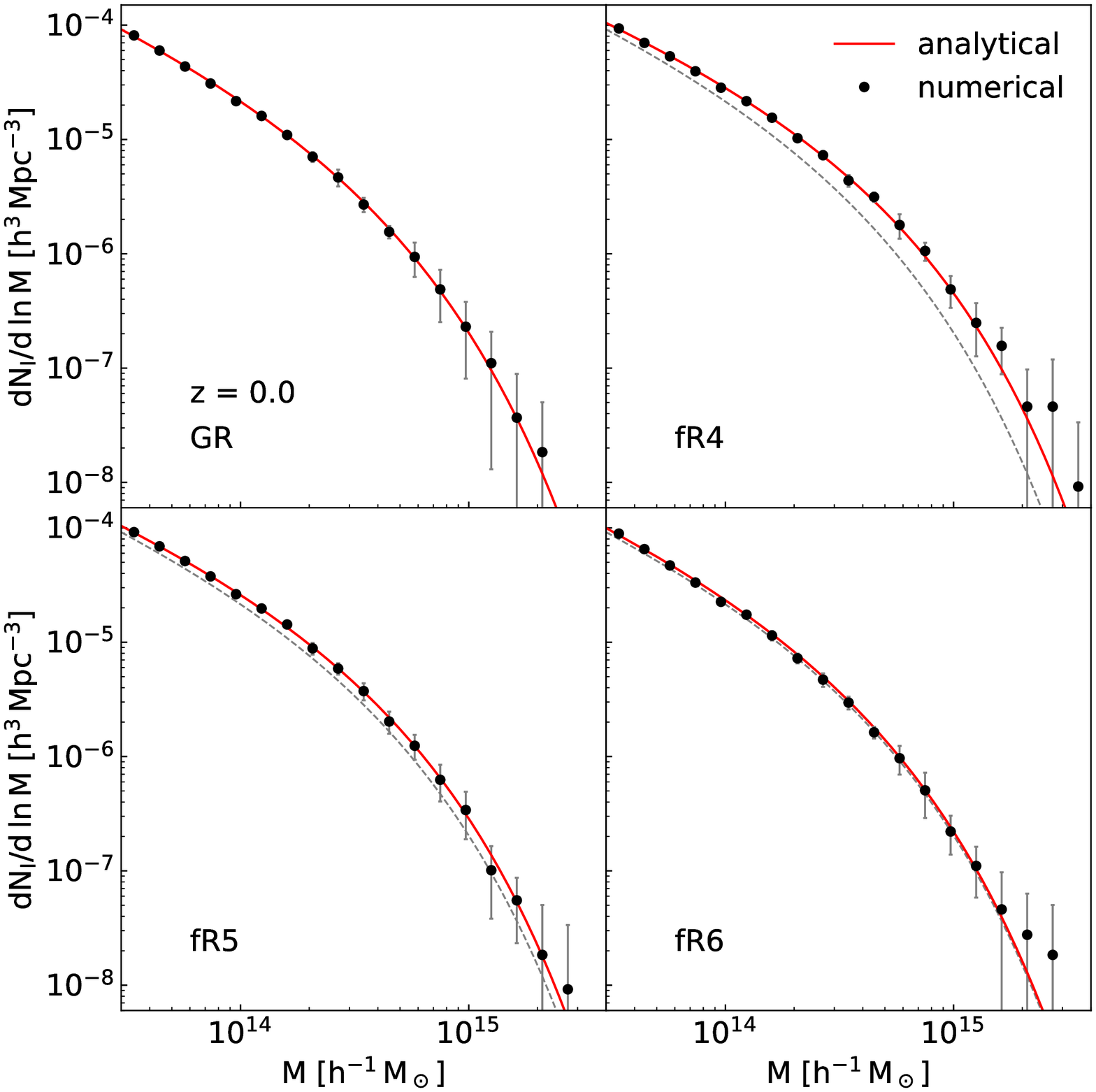}
\caption{Same as Figure \ref{fig:mf_cde_z0} but for four different $f(R)$ gravity cosmologies.}
\label{fig:mf_fr_z0}
\end{center}
\end{figure}

\begin{figure}[ht]
\begin{center}
\plotone{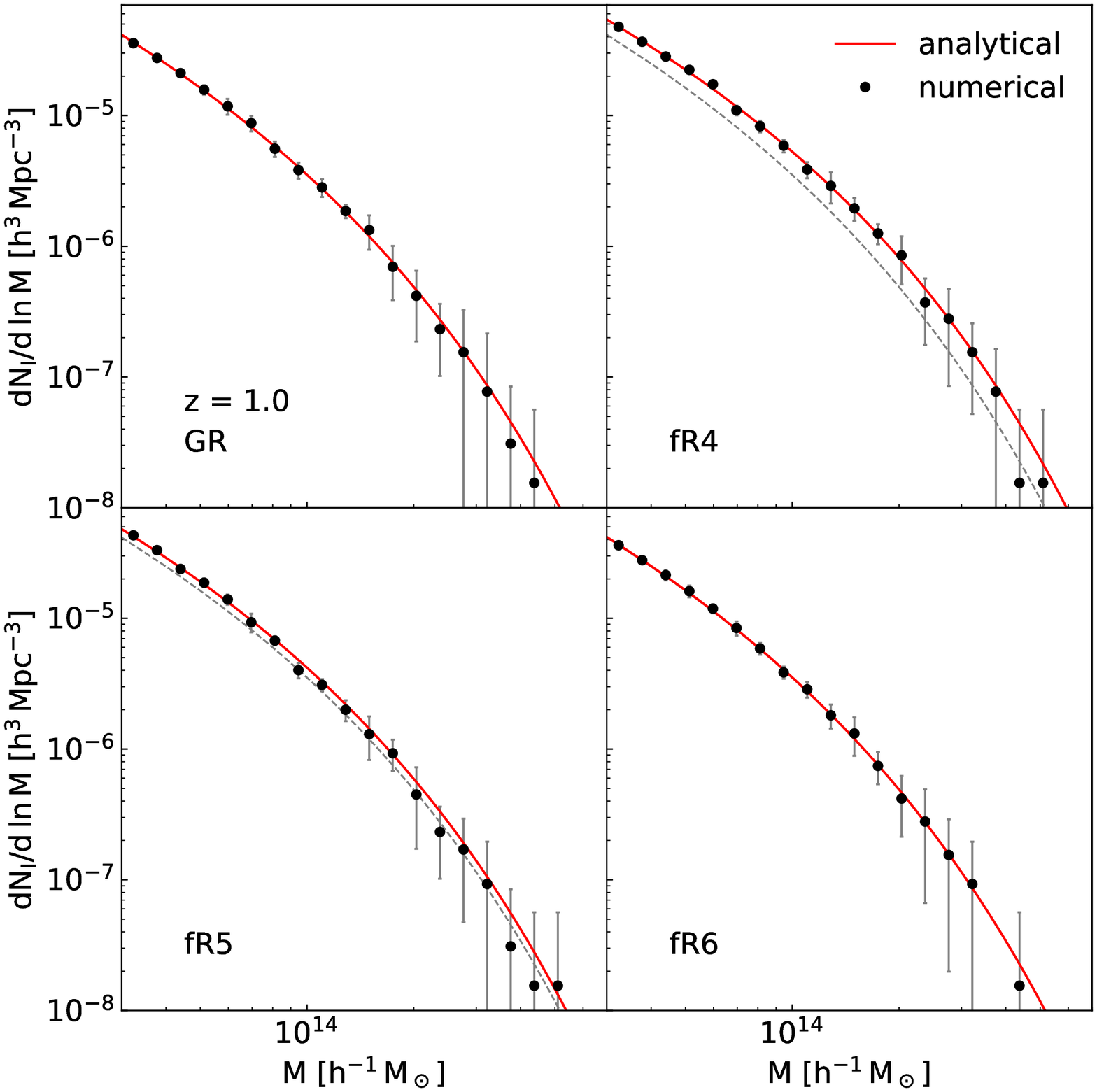}
\caption{Same as Figure \ref{fig:mf_fr_z0} but at $z=1$.}
\label{fig:mf_fr_z1}
\end{center}
\end{figure}

\begin{figure}[ht]
\begin{center}
\plotone{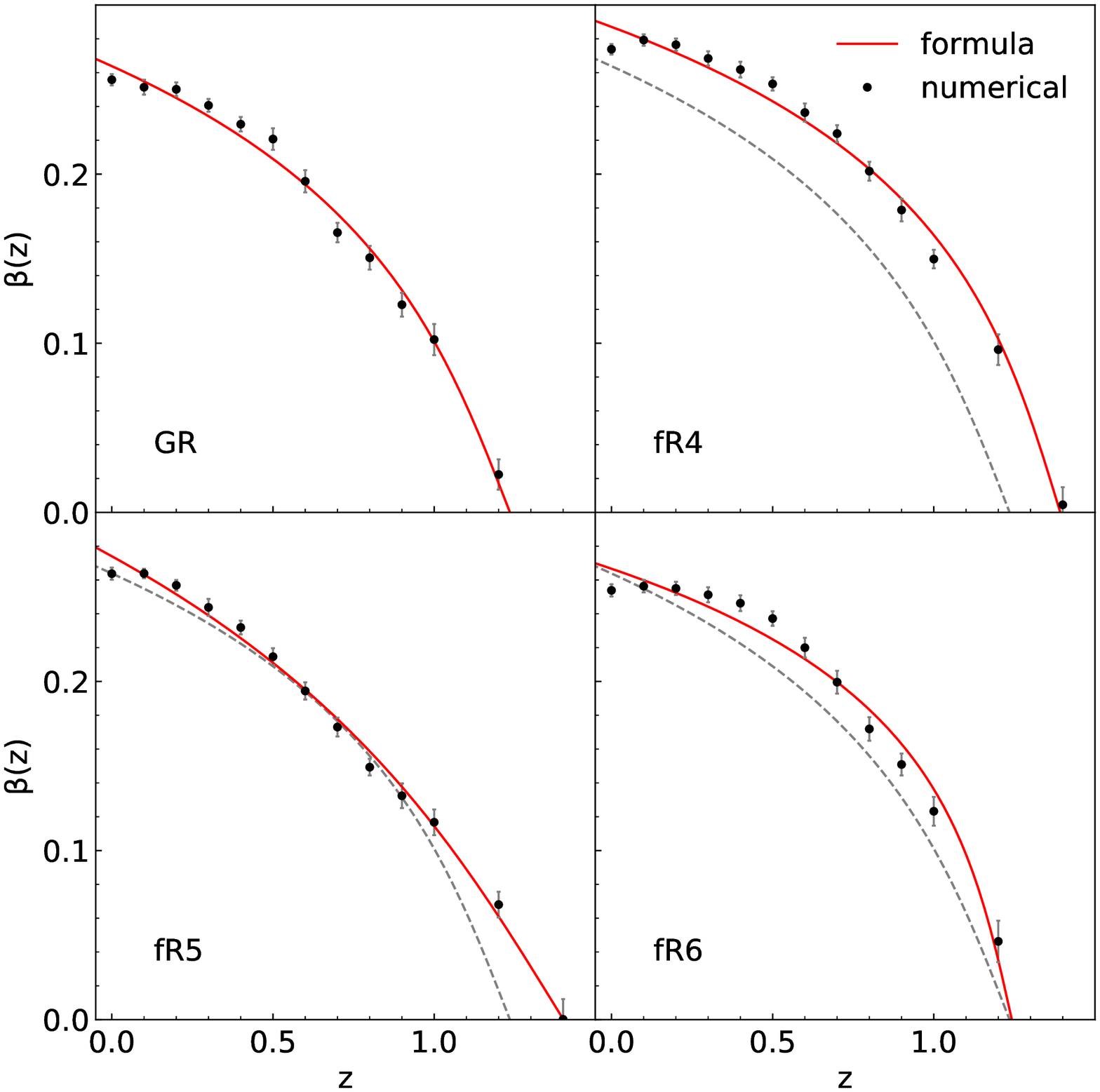}
\caption{Same as Figure \ref{fig:beta_cde} but for the $f(R)$ cosmologies.}
\label{fig:beta_fr}
\end{center}
\end{figure}

Figure \ref{fig:mf_fr_z0} (Figure \ref{fig:mf_fr_z1}) depicts the same as Figure \ref{fig:mf_cde_z0} (Figure \ref{fig:mf_cde_z1}) but for the fR4, fR5 and fR6 cases, 
revealing that the analytic model matches quite well the numerical results even for the $f(R)$ gravity models. As expected, the fR4 (fR6) yields the most (least) abundant field 
clusters in the entire mass range. No statistically significant difference is found in $dN_{\rm I}/d\ln M$ between the GR and the fR6 cases, indicating their mutual degeneracy 
in the field cluster mass functions. Figure \ref{fig:beta_fr} plots the same as Figure \ref{fig:beta_cde} but for the fR4, fR5 and fR6 cases. As can be seen, despite that the 
field cluster mass functions fail to distinguish between the GR and the fR6 cases, the field cluster drifting coefficient, $\beta(z)$, can break the degeneracy, showing a substantial 
difference between the two cosmologies. 

It is worth noting the distinct redshift dependence of the difference in $\beta(z)$  between the GR and each $f(R)$ gravity cosmology.  
The fR4 case yields an almost redshift-independent shift of  $\beta(z)$ from that of the GR case, while the other two cases show different redshift-dependent shifts 
between each other. That is, for the fR5 (fR6) case, the largest deviation of $\beta(z)$ from that of the GR case occurs at the low (high) redshift ends. 
A qualitative explanation for this trend is provided in the following.
For the fR4 case, the fifth-force is basically always unscreened at low redshifts, which implies that the haloes of all masses are equally affected by the 
fifth force, and that there is no sharp transition between screened and unscreened regions in the cosmic web. 

Whereas, for the fR5 and fR6 cases, as the screening properties imply that the massive halos are screened, while less massive halos are not. 
This introduces a mass dependence in the deviation of the halo mass function from that of the GR case. In particular, for the fR6 case, the high-mass tail of the halo mass function 
should be almost unaffected and thus there should be an enhancement in the number of smaller-mass halos. This may have a different impact on the bias of halos at different 
masses, and consequently an impact on the definition of the field clusters (i.e. isolated massive objects) which could induce a different evolution of $\beta(z)$.
A more quantitative analysis is required to fully understand the distinct differences in $\beta(z)$ between the GR and each $f(R)$ cosmology, 
which is, however, beyond the scope of this paper.

\subsection{Combined Effect of $f(R)$+$\nu$ on $\beta(z)$}\label{sec:frnu}

\begin{figure}[ht]
\begin{center}
\plotone{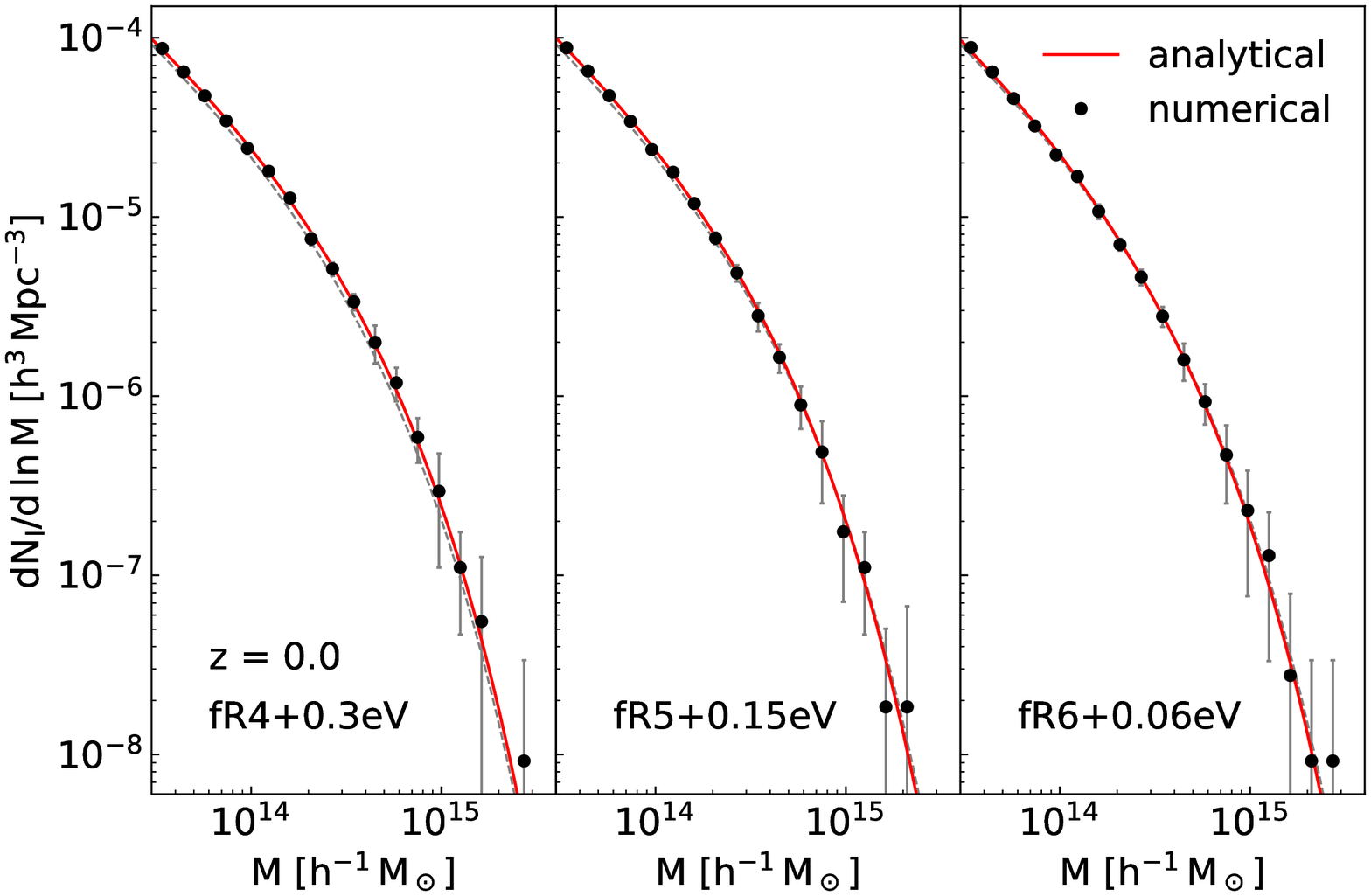}
\caption{Same as Figure \ref{fig:mf_cde_z0} but for four different $f(R)$ gravity+$\nu$ cosmologies.}
\label{fig:mf_frnu_z0}
\end{center}
\end{figure}

\begin{figure}[ht]
\begin{center}
\plotone{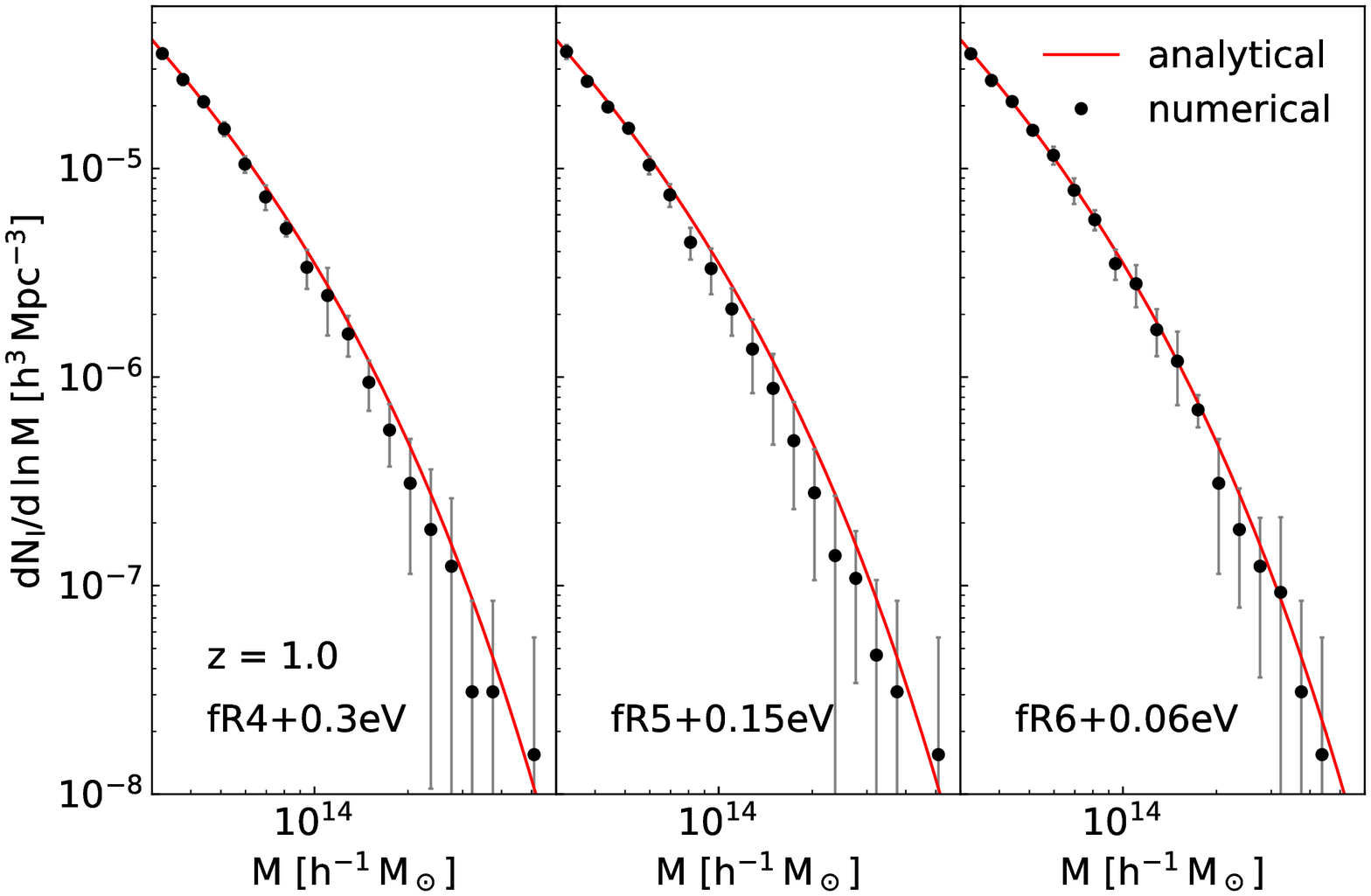}
\caption{Same as Figure \ref{fig:mf_frnu_z0} at $z=1$.}
\label{fig:mf_frnu_z1}
\end{center}
\end{figure}

\begin{figure}[ht]
\begin{center}
\plotone{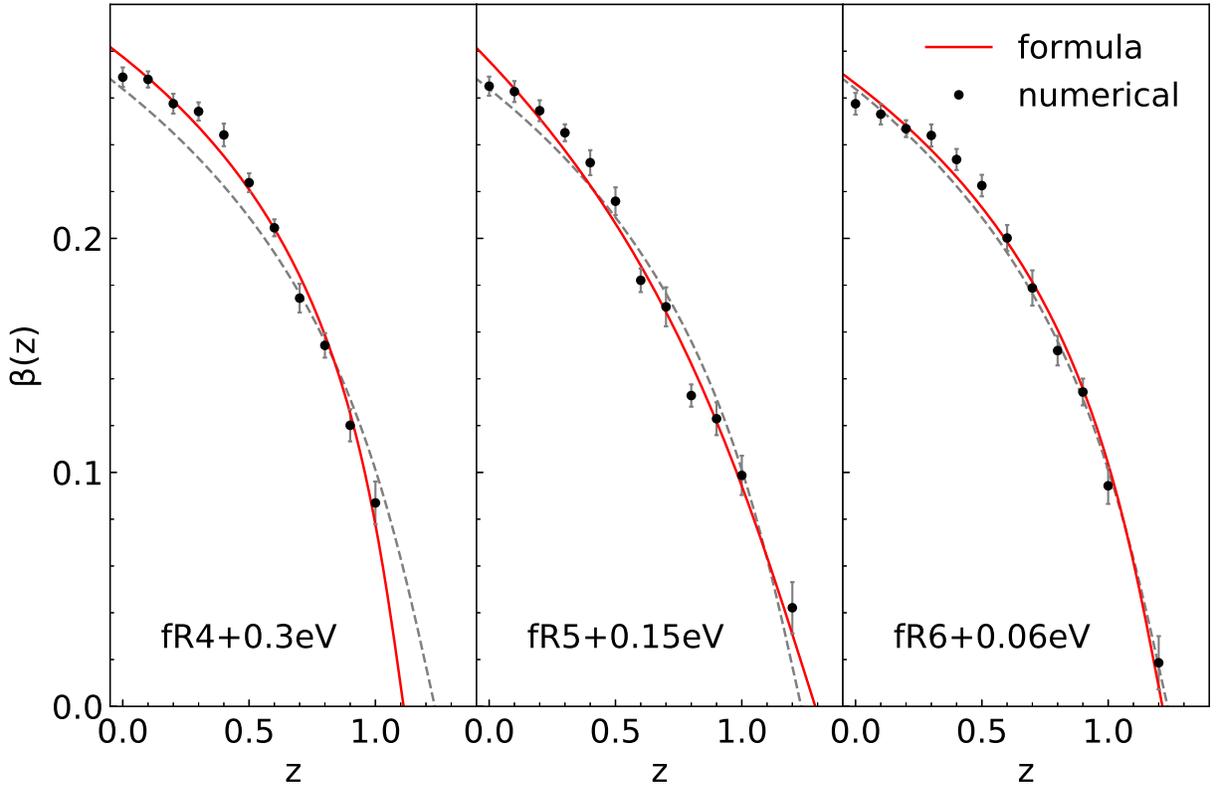}
\caption{Same as Figure \ref{fig:beta_cde} but for the $f(R)$ gravity + $\nu$ cosmologies.}
\label{fig:beta_frnu}
\end{center}
\end{figure}

Now that the validity of the analytic model of the field cluster mass function for the $f(R)$ gravity cosmology is confirmed, we repeat the whole 
process but for the fR4+$0.3\ev$, fR5+$0.15\ev$ and fR6+$0.06\ev$ cosmologies, which were shown to be degenerate with the GR in the standard statistics 
including the nonlinear density power spectrum, cluster mass functions and halo bias \citep{bal-etal14,gio-etal19}. 
Figure \ref{fig:mf_frnu_z0} (Figure \ref{fig:mf_frnu_z1}) depicts the same as Figure \ref{fig:mf_cde_z0} (Figure \ref{fig:mf_cde_z1}) but for the 
fR4+$0.3\ev$, fR5+$0.15\ev$ and fR6+$0.06\ev$ cosmologies. As can be seen, the analytic model is still quite valid in matching the numerically obtained field cluster 
mass functions even for the $f(R)$+$\nu$ cosmologies. At $z=0$, the three $f(R)$+$\nu$ cosmologies show no difference from the GR case in the field cluster mass 
functions. At $z=1$, the differences in $dN_{\rm I}/d\ln M$ between the $f(R)$+$\nu$ and the GR cases are slightly larger but still not statistically significant. 
Figure \ref{fig:beta_frnu} plots the same as Figure \ref{fig:beta_cde} but for the fR4+$0.3\ev$, fR5+$0.15\ev$ and fR6+$0.06\ev$ cosmologies.
As can be seen, the fR4+$0.3\ev$ cosmology yields a substantial difference in $\beta(z)$ from the GR case, in spite of their mutual degeneracy 
in the standard statistics. Yet, both of the fR5+$0.15\ev$ and the fR6+$0.06\ev$ cosmologies show almost no difference in $\beta(z)$ from the GR case. 

\begin{figure}[ht]
\begin{center}
\plotone{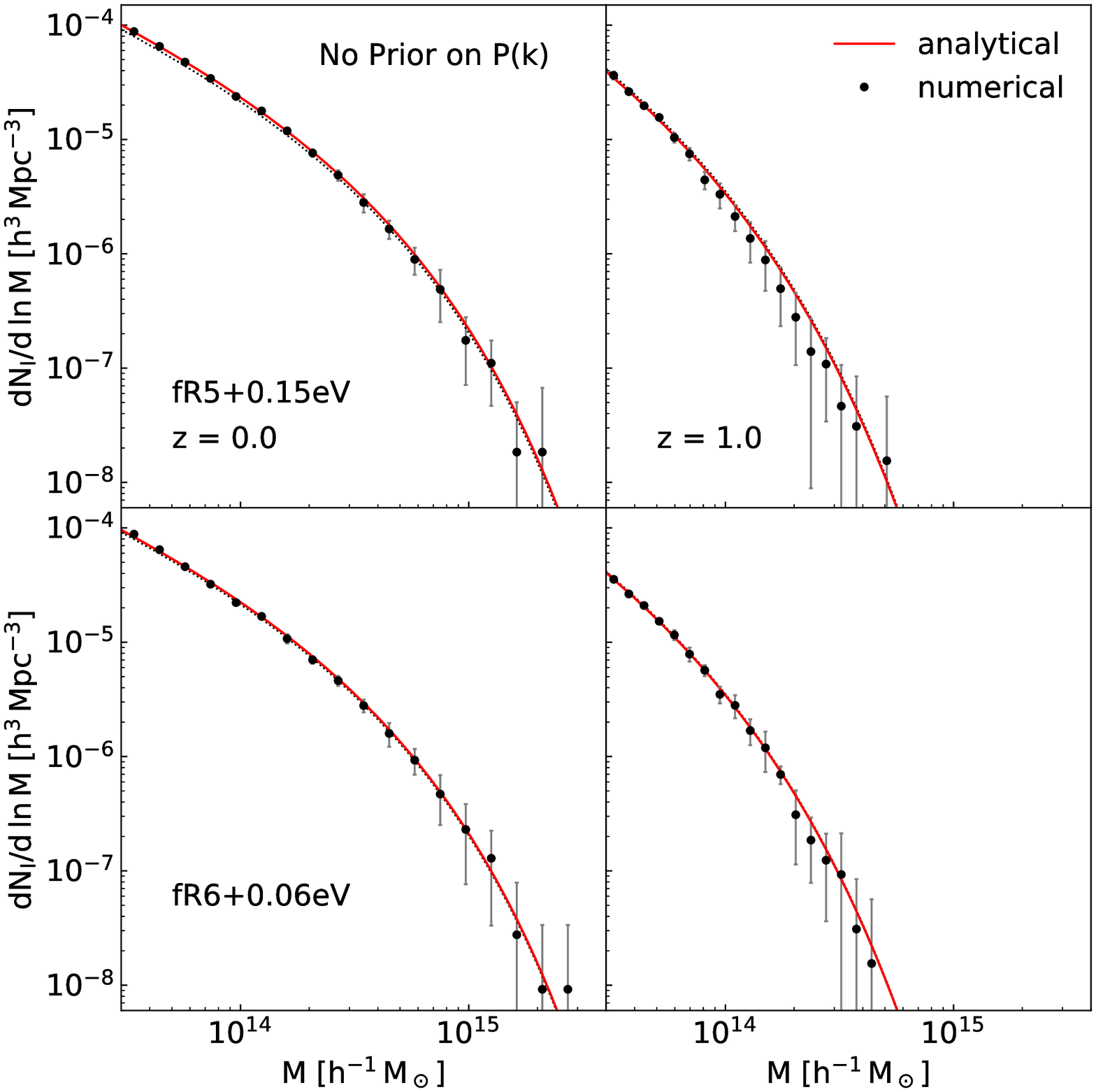}
\caption{Field cluster mass functions for the fR5+$0.1\ev$ and fR6+$0.05\ev$ cases determined without 
using prior information on $P(k)$.}
\label{fig:noprior_mf_frnu}
\end{center}
\end{figure}

As done in Section \ref{sec:cde}, we redetermine $dN_{\rm I}/d\ln M$ for both of the fR5+$0.15\ev$ and fR6+$0.06\ev$ cases without using prior information 
on the shapes of their power spectra, which are plotted in Figure \ref{fig:noprior_mf_frnu}. As can be seen, the analytic model, Equation (\ref{eqn:exc_f}), still agrees quite 
well with the numerically obtained field cluster mass functions for both of the cases at both of the redshifts, despite that $P(k;GR)$ is substituted for $P(k;{\rm fR5}+0.15\ev)$ 
and $P(k;{\rm fR6}+0.06\ev)$. The drifting coefficient, $\beta(z)$, redetermined without using prior information is plotted in Figure \ref{fig:noprior_beta_frnu}, which reveals 
that the three cosmologies yield much larger differences in $\beta(z)$. 

\begin{figure}[ht]
\begin{center}
\plotone{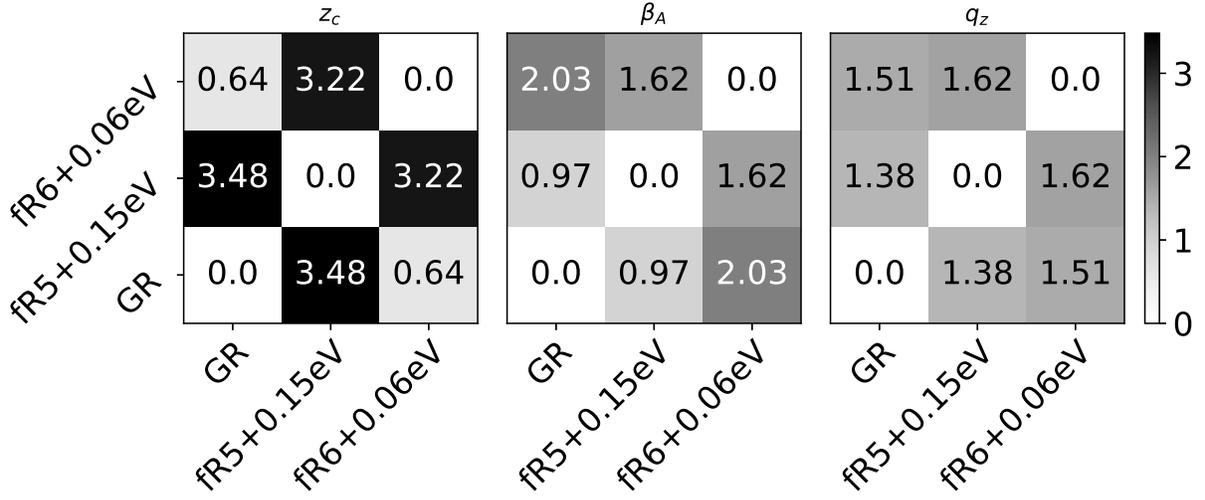}
\caption{Statistical significances of the differences in $z_{c}, \beta_{A}$ and $q_{z}$ among the GR, fR5+$0.1\ev$ and fR6+$0.05\ev$ cases 
that are mutually degenerate in the standard diagnostics.}
\label{fig:signi_frnu}
\end{center}
\end{figure}

For each cosmology, we determine the best-fit values of $\beta_{A},\ q_{z}$ and $z_{c}$ by fitting Equation (\ref{eqn:inv_sinh}) to $\beta(z)$ obtained without priors. Then, we calculate 
the statistical significance of the differences in the three parameters among the three cosmologies, which are shown in Figure \ref{fig:signi_frnu}. As can be seen, without using prior information 
on the linear density power spectra of the $f(R)$+$\nu$ cosmologies, the statistical significance of the differences in $z_{c}$ between the GR and the fR5+0.15$\ev$ and between the 
fR6+$0.06\ev$ and the fR5+$0.15\ev$ are as high as $3.48$ and $3.22$, respectively.

Meanwhile, for the fR6+0.06$\ev$ case, it turns out to be not $z_{c}$ but $\beta_{A}$ that is able to distinguish it from the GR case with $\Delta \beta_{A}/\sigma_{\Delta \beta_{A}}=2.03$. The lower statistical significance of the differences in $\beta(z)$ between the GR and the fR6+0.06$\ev$ is likely to be at least partially due to the large errors caused by the relatively small 
box size of the DUSTGRAIN-pathfinder simulations. Given the distinct behaviors of $\beta(z)$ between the the GR and the fR6+0.06$\ev$ shown in Figure \ref{fig:noprior_beta_frnu}, 
we suspect that if a halo sample from a larger simulations were used, the statistical significance would increase.
The best-fit values of $\beta_{A},\ q_{z}$ and $z_{c}$ for each of the seven cosmologies simulated by the DUSTGRAIN-pathfinder are shown in Table \ref{tab:mds_frnu}. 
For the fR6+0.06$\ev$ and fR5+0.15$\ev$ cosmologies that are degenerate with the GR case in the standard statistics, what is listed in Table \ref{tab:mds_frnu} is the best-fit values 
obtained without using priors in the shapes of the linear density power spectra.

\section{Summary and Discussions}\label{sec:con}

The new diagnostic developed by \citet{RL20a} traces the redshift evolution of the drifting coefficient, $\beta(z)$, which is a single parameter of the analytic model 
for the field cluster mass function derived by \citet{lee12} in the frame of the generalized excursion set theory \citep{MR10a,MR10b,CA11a,CA11b}.
Motivated by the recent finding that this diagnostic can in principle break the $\sigma_{8}$-$\sum m_{\nu}$ degeneracy \citep{RL20b}, 
we have studied whether or not the same diagnostic can break the degeneracy between the non-standard and the standard $\Lambda$CDM+GR cosmologies by 
utilizing the data from the CoDECS and DUSTGRAIN-pathfinder simulations. 
For this study, we have considered eleven different non-standard cosmologies which include $5$ different cDE 
(EXP001, EXP002, EXP003, EXP008e3 and SUGRA), $3$ different $f(R)$ gravity (fR4, fR5, fR6), and $3$ different $f(R)$ gravity+$\nu$ cosmologies 
(fR4+0.3eV, fR5+0.15eV, fR6+0.06eV). 

Among the cDE and $f(R)$ gravity cosmologies, the EXP001 and fR6 have been known to be very similar to the $\Lambda$CDM+GR in the linear density 
power spectra and cluster mass functions at $z=0$, due to their extremely weak DM-DE coupling and fifth force, respectively. 
The three $f(R)$ gravity+$\nu$ cosmologies have been known to be degenerate not only with the $\Lambda$CDM+GR but also among one another in the 
standard diagnostics that include the cluster mass functions, halo bias, and nonlinear density power spectrum \citep{bal-etal14,gio-etal19}.
Analyzing the catalogs of the FoF bound objects for each cosmology, we have identified the field cluster halos and determined their mass functions at each redshift in the range 
of $0\le z\le 1$. The best-fit value of the drifting coefficient $\beta$ has been found by adjusting the analytic model of \citet{lee12} to the numerically determined field cluster 
mass functions. The fitting formula for $\beta(z)$ proposed by \citet{RL20a} have been used to assess the statistical significance of the 
differences in $\beta(z)$ among the cosmologies. This analysis has lead us to find the following.
\begin{itemize}
\item 
The analytic model of \citet{lee12} for the field cluster mass functions with the best-fit values of $\beta$ agrees excellently well with the numerical results at all redshifts for 
all of the non-standard cosmologies. 
\item
The empirical formula of \citet{RL20a} for $\beta(z)$ works fairly well for all of the non-standard cosmologies. 
\item
Despite that they produce very similar (field) cluster mass functions, the $\Lambda$CDM and the SUGRA cosmologies substantially differ in $\beta(z)$ from each other. 
\item
The degeneracy between the $\Lambda$CDM and the EXP001 in the (field) cluster mass functions can be broken by $\beta(z)$ with $2.53\sigma$ significance 
without using any prior information on the linear density power spectrum, $P(k;{\rm EXP001})$. 
\item
The degeneracy between the $\Lambda$CDM+GR and the fR4+0.3eV in the linear density power spectra and (field) cluster mass functions can be broken by 
$\beta(z)$ with high statistical significance. 
\item
The degeneracy among the $\Lambda$CDM+GR, fR5+0.15eV and fR6+0.05eV cosmologies in the standard diagnostics can be broken by $\beta(z)$ with minimum 
$2.01\sigma$ significance, without using any prior information on the linear density power spectra. 
\end{itemize}

To understand the advantage of using $\beta(z)$ as a cosmology discriminator, it may be worth comparing $\beta(z)$ with the standard diagnostics such as the linear density 
power spectrum, nonlinear density bi spectrum and cluster mass function.
As for the linear density power spectrum, it deals with isotropically averaged densities and thus fail to capture independent information contained in the anisotropic 
nonlinear cosmic web about the background cosmology \citep{nai-etal20}. 
As for the nonlinear density bi spectrum that treats the nonlinear anisotropic density field, it is not readily observable, suffering from 
highly nonlinear halo bias and redshift space distortion effects. 
Regarding the cluster mass function, although it is free from the halo bias and redshift space distortion effect, it varies most sensitively with the value of $\sigma_{8}$.
If two different cosmologies share an identical value of $\sigma_{8}$ (e.g., $\Lambda$CDM and SUGRA), the cluster mass function is apt to fail in telling them apart. 

Meanwhile, the field cluster drifting coefficient, $\beta(z)$, deals with the non-spherical collapse occurring in the anisotropic cosmic web that contains additional information on the 
initial conditions. It is free from the halo bias and redshift space distortion effect, directly quantifying how the background cosmology deviates from the Einstein-de Sitter state which 
sensitively depends on the evolution of the energy contents of the universe.
Notwithstanding, we have yet to find a direct link of $\beta(z)$ to the initial conditions, which weakens its power as a probe of gravity and dark sector physics.
The very fact that the inverse sine hyperbolic function provides a fairly good approximation to the empirically determined $\beta(z)$ for all of the cosmologies hints that 
it should be beyond a mere fitting formula. Our future work is in the direction of theoretically deriving $\beta(z)$ from first principles, providing a physical explanation for 
why $\beta(z)$ behaves as an inverse sine hyperbolic function of $z$ and establishing its direct link to the initial conditions. 

\begin{figure}[ht]
\begin{center}
\plotone{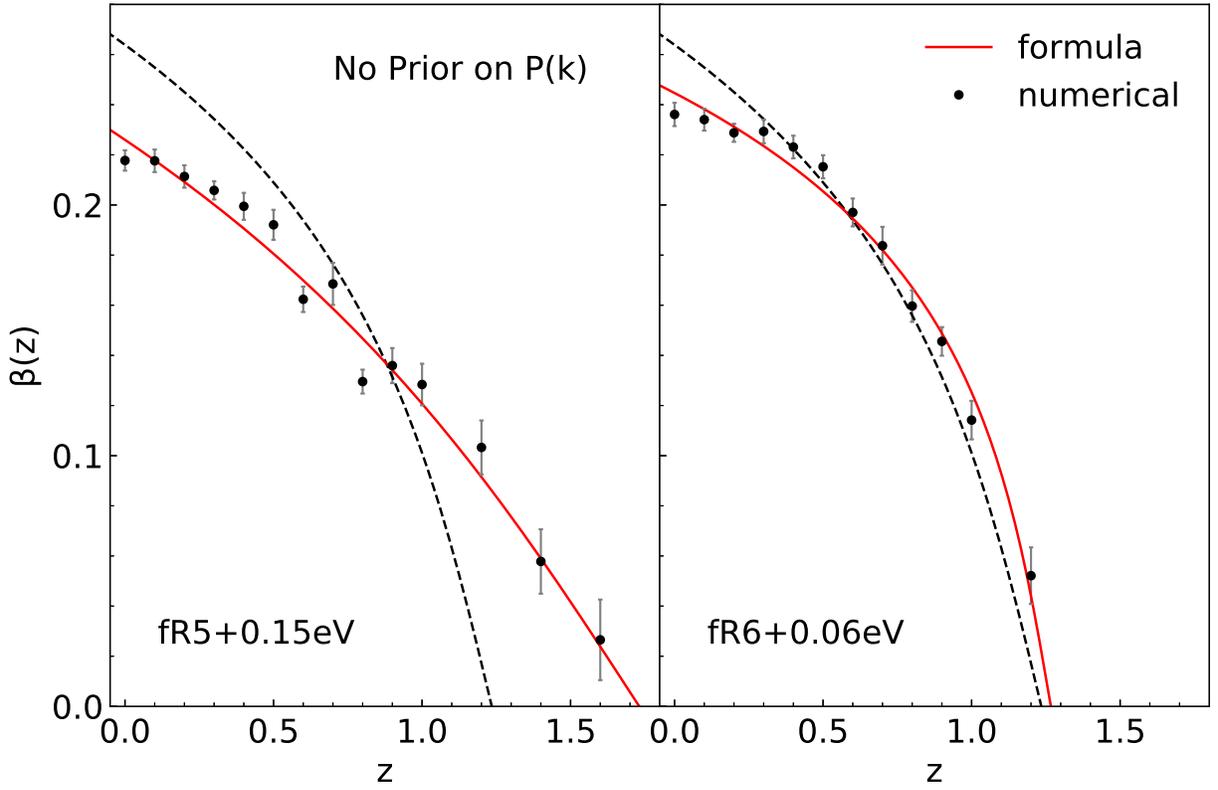}
\caption{$\beta(z)$ for the fR5+$0.1\ev$ and fR6+$0.05\ev$ cases determined without 
using prior information on $P(k)$.}
\label{fig:noprior_beta_frnu}
\end{center}
\end{figure}

Another advantage is high observational applicability of $\beta(z)$. As shown in Figure \ref{fig:noprior_beta_frnu},  the fR5+0.15eV and fR6+0.06eV cases substantially differ 
from the GR case in the low-$z$ values of $\beta(z)$ ($z<0.5$). In other words, it does not require a large sample of the high-$z$ clusters with $z>0.5$ to distinguish between the 
$f(R)$+$\nu$ and the GR cases with $\beta(z)$ in practice.  
Yet, to distinguish between the fR5+0.15eV and the fR6+0.06eV cases as well as between the $f(R)$ gravity and the cDE cosmologies with $\beta(z)$, it indeed requires  
a large cluster sample from a wide range of redshifts.  The upcoming large-scale deep surveys such as the LSST (Large Synoptic Survey Telescope) or EUCLID 
that is expected to cover the redshift range up to $z\sim 2$ \citep{lsst,euclid} will improve prospects for $\beta(z)$ as a cosmological discriminator. 

\acknowledgments

We thank an anonymous referee for useful comments and suggestions. 
JL acknowledges the support by Basic Science Research Program through the National Research Foundation (NRF) of Korea 
funded by the Ministry of Education (No.2019R1A2C1083855). 
MB acknowledges support by the project "Combining Cosmic Microwave Background and Large Scale Structure data: an Integrated Approach for Addressing Fundamental Questions in Cosmology", funded by the MIUR Progetti di Ricerca di Rilevante Interesse Nazionale (PRIN) Bando 2017 - grant 2017YJYZAH. The N-body simulations described in this work have been performed on the Hydra supercomputer at RZG and on the Marconi supercomputer at Cineca thanks to the PRACE allocation 2016153604 (P.I. M. Baldi).

\clearpage
\clearpage
\clearpage
\clearpage
\clearpage
\clearpage
\clearpage
\clearpage
\clearpage
\clearpage
\clearpage
\clearpage
\clearpage
\clearpage
\clearpage

\clearpage
\begin{deluxetable}{cccccccc}
\tablewidth{0pt}
\tablecaption{Best-fit Parameters of $\beta(z)$ for the CoDECS cosmologies.}
\setlength{\tabcolsep}{2mm}
\tablehead{Model & $V(\phi)$ & $s$ & $\sigma_8$ & $\beta_A$ & $q_z$ & $z_c$}
\startdata
$\Lambda$CDM		& -												& - 			& 0.809 & $-0.16\pm0.01$ & $0.31\pm0.04$ & $1.10\pm0.02$ \\
EXP001					& $e^{-0.08\phi}$ 					& 0.05 	& 0.825 & $-0.16\pm0.01$ & $0.31\pm0.05$ & $1.04\pm0.02$ \\
EXP002					& $e^{-0.08\phi}$ 					& 0.10		& 0.875 & $-0.17\pm0.02$ & $0.35\pm0.08$ & $1.32\pm0.04$ \\
EXP003					& $e^{-0.08\phi}$ 					& 0.15 	& 0.967 & $-0.14\pm0.01$ & $0.19\pm0.06$ & $1.44\pm0.05$ \\
EXP008e3				& $e^{-0.08\phi}$ 					& 0.40		& 0.895 & $-0.16\pm0.01$ & $0.27\pm0.04$ & $1.19\pm0.03$ \\
SUGRA003				& $\phi^{-2.15}e^{\phi^2/2}$	& -0.15 	& 0.806 & $-0.16\pm0.01$ & $0.39\pm0.07$ & $1.35\pm0.03$ 
\enddata
\label{tab:mds_cde}
\end{deluxetable}
\clearpage
\begin{deluxetable}{ccccccc}
\tablewidth{0pt}
\tablecaption{Best-fit Parameters of $\beta(z)$ for the DUSTGRAIN-pathfinder cosmologies.}
\setlength{\tabcolsep}{3mm}
\tablehead{Model & $|f_{R0}|$ & $\sum m_{\nu}\,[{\rm eV}]$ & $\sigma_8$ & $\beta_A$ & $q_z$ & $z_c$}
\startdata
$\Lambda$CDM		& - 					& 0.0		& 0.847 & $-0.11\pm0.01$ & $0.22\pm0.06$ & $1.24\pm0.03$ \\
fR4							& $10^{-4}$	& 0.0		& 0.967 & $-0.10\pm0.01$ & $0.16\pm0.04$ & $1.39\pm0.03$ \\
fR5							& $10^{-5}$	& 0.0 		& 0.903 & $-0.16\pm0.02$ & $0.50\pm0.11$ & $1.40\pm0.04$ \\
fR6							& $10^{-6}$	& 0.0 		& 0.861 & $-0.08\pm0.01$ & $0.09\pm0.04$ & $1.24\pm0.04$ \\
fR4+0.3eV				& $10^{-4}$	& 0.3 		& 0.893 & $-0.09\pm0.01$ & $0.29\pm0.09$ & $1.52\pm0.06$ \\
fR5+0.15eV			& $10^{-5}$	& 0.15		& 0.864 & $-0.15\pm0.05$ & $0.85\pm0.45$ & $1.73\pm0.14$ \\
fR6+0.06eV			& $10^{-6}$	& 0.06		& 0.847 & $-0.08\pm0.01$ & $0.11\pm0.04$ & $1.27\pm0.04$
\enddata
\label{tab:mds_frnu}
\end{deluxetable}

\end{document}